# Agentic AI and Multiagentic: Are We Reinventing the Wheel?

Vicent Botti

Valencian Research Institute for Artificial Intelligence (VRAIN)

Universitat Politècnica de València (UPV)

Valencian Graduated School and Research Network of Artificial Intelligence (ValgrAI)

*vbotti@vrain.upv.es*

## Abstract

The terms "**Agentic AI**" and "**Multiagentic AI**" have recently gained popularity in discussions on generative artificial intelligence, often used to describe autonomous software agents and systems composed of such agents. However, the use of these terms confuses these buzzwords with well-established concepts in AI literature: **intelligent agents** and **multi-agent systems**. This article offers a critical analysis of this conceptual misuse. We review the theoretical origins of "agentic" in the social sciences (Bandura, 1986) and philosophical notions of intentionality (Dennett, 1971), and then summarise foundational works on intelligent agents and multi-agent systems by Wooldridge, Jennings and others. We examine classic agent architectures—from simple reactive agents to Belief-Desire-Intention (BDI) models—and highlight key properties (autonomy, reactivity, proactivity, social capability) that define agency in AI. We then discuss recent developments in large language models (LLMs) and agent platforms based on LLMs, including the emergence of LLM-powered AI agents and open-source multi-agent orchestration frameworks. We argue that the term "AI Agentic" is often used as a buzzword for what are essentially AI agents, and "AI Multiagentic" for what are multi-agent systems. This confusion overlooks decades of research in the field of autonomous agents / multi-agent systems. The article advocates for scientific and technological rigour and the use of established terminology from the state of the art in AI, incorporating the wealth of existing knowledge—including standards for multi-agent system platforms, communication languages and coordination/cooperation algorithms, agreement technologies (automated negotiation, argumentation, virtual organisations, trust, reputation, etc.)—into the new and promising wave of LLM-based AI agents, so as not to end up **reinventing the wheel**.

## Introduction

Recent advances in **generative AI**, particularly large language models (LLMs), have led to a resurgence of interest in **autonomous software agents**: AI systems that can autonomously perceive their environment, reason, and act to meet their design goals to perform tasks on behalf of users. Visionaries such as Bill Gates predict that, in the near future, we will all have a personal AI assistant "far superiror to current technology," capable of responding to natural language requests and performing various tasks by understanding the user's goals [6]. Gates notes that this software, which he and others have imagined for decades, is finally becoming practical thanks to advances in AI, and that such "**agents**" could revolutionize our interaction with computers. In parallel, some industrial discourse, not always; there are many cases where the terminology is respected, has introduced new terms such as "**Agentic AI**" to describe AI systems endowed with autonomy and proactive decision-making. Technical blogs and articles differentiate between "AI agents" and "Agentic AI," presenting the latter as a general **framework** where multiple agents operate. Similarly, the term "**Multiagentic**" is sometimes used for systems with multiple interacting agents, analogous to classic **multiagent systems**.

This emergence of "agentic" terminology in AI raises critical questions. Are these genuinely new concepts, or simply new labels for already established ideas in autonomous agents research? The field of **intelligent agents and multi-agent systems** (MAS) has a rich history dating back decades, with its own theoretical foundations, architectures, and even dedicated conferences, journals and scientific associations of researchers in the field. Before embracing the hype of "agentic AI," it is important to examine whether the term is being misapplied as a buzzword for capabilities long understood in AI – In other words, are we **reinventing the wheel** by equating *agentic AI* with *intelligent agents* and *multiagentic systems* with *multi-agent systems*?

In this article, we undertake a comprehensive analysis of the concept of agency in AI. We begin by revisiting the origins of the term *agentic* in social cognition and the philosophical basis for treating systems as intentional agents. Next, we summarize foundational work in AI on intelligent agents and MAS, including core definitions and properties, and the evolution of agent architectures (reactive, deliberative, hybrid, BDI, etc.). We then discuss recent trends in LLM-based autonomous agents – sometimes labeled as agentic AI – and the proliferation of frameworks for "agentic" systems. Throughout, we highlight how the new usage of *agentic* and *multiagentic* often maps directly onto classical concepts. Finally, we argue for precise terminology, with scientific rigor, and for leveraging the extensive existing research and its results in the area of agents/MAS to build the next generation of AI systems where generative AI plays a central role, rather than ignoring previous work. By doing so, the AI community can avoid conceptual confusion and truly advance the state of the art in autonomous agents. Scientific rigor requires a thorough study of the state of the art in any discipline before beginning work in it.

## The Origins of "Agentic": Agency in Psychology and Philosophy

The adjective "**agentic**" gained formal prominence outside of computer science. The first reference to the term "**agentic**" appears in the context of social psychology, in the work of psychologists such as **Richard H. Goffman and Susan Fiske** in the 1970s, although its roots come from earlier concepts related to agency and autonomy. The term comes from "**agency**" and refers to the capacity of individuals to act autonomously, make decisions, and exercise control over their lives. The concept of agency has existed for much longer, but the term "agentic" as an adjective was popularized in academic literature especially through the work of **Albert Bandura**. In his book Social Foundations of Thought and Action: A Social Cognitive Theory (1986), Albert Bandura introduced the term agentic in the context of human agency [1]. In Bandura's social cognitive theory, agentic refers to the capacity of individuals to act intentionally, make decisions, and exercise control over their environment. This emphasizes a proactive and deliberate aspect of human behavior: in short, the quality of having agency. A person with agentic capacity doesn't just react to external forces; they are originators of actions, pursuing goals of their own free will. Bandura's introduction of the term highlighted the proactive and self-regulatory nature of human action, underscoring that people are agents of their own change. This concept of agency has been influential in psychology, underpinning theories of self-efficacy and motivation, but it was originally applied to humans, not machines.

If we look in the dictionary we find the following meaning of agentic:

> *"Agentic" is an adjective derived from the noun agency, and it's commonly used in psychology, sociology, and education. It refers to having the capacity to act independently, make choices, and exert control over one's life and circumstances.*
> *Definition (in context):*
> > *Exhibiting agency; capable of intentional action and decision-making.*
> > *Acting as an agent (i.e., initiating and directing one's behavior).*

In parallel, philosophers such as Daniel C. Dennett were exploring how attributions of agency and intentionality could be applied to machines and other non-human systems. Dennett's 1971 paper, "**Intentional Systems**," articulated the idea that we often interpret and predict the behavior of complex systems by treating them as if they had beliefs, desires, and intentions [2]. Dennett called an intentional system one whose behavior can be successfully explained and predicted by attributing mentalistic qualities like beliefs and desires to it, similar to the folk psychology we use for humans [2]. This provides a philosophical basis for talking about **AI agents**: even if a thermostat or a program does not literally have desires or beliefs, treating it as an intentional agent (with goals, knowledge, etc.) can be a useful abstraction for understanding its behavior [2]. Dennett's theory thus bridged the gap between human psychology and artificial entities, suggesting that any system whose behavior is sufficiently complex and directed can be viewed in intentional terms [2]. Notably, this coincided with early research in AI, robotics, and cybernetics; it provided a justification for describing machines with language usually reserved for conscious agents.

The notions from Bandura and Dennett provide a backdrop for *agency* as a concept: **agency** implies intentional, goal-directed action, whether in humans or artificial systems. When we call an AI system "agentic," in the strict sense we mean it exhibits agency – i.e. it can **act autonomously with intention** to achieve objectives. In human psychology this is uncontroversial, but in AI it raises the question: what does it take for an artificial system to be considered an *agent*? The field of AI has answered this over the years by defining the intelligent agent abstraction, which we examine next. Suffice to say, by the time the term *agentic* started being applied to AI in the 2020s, the community already had a deep understanding of what it means for a system to have agency.

## Intelligent Agents: Definitions and Key Properties

In 1991, Yoav Shoham proposed a new programming paradigm, 'Agent-oriented Programming' [16]. The paradigm promotes a social view of computing, in which multiple "agents" interact with each other.

In artificial intelligence, an **intelligent agent** is fundamentally an autonomous entity that perceives its environment and acts upon it to achieve its goals. A classic definition by **Wooldridge and Jennings** (1995) [3] describes an intelligent agent as "*a computing system situated in an environment, capable of flexible and autonomous action to meet its design objectives*" (Fig. 1). This encompasses several important aspects. The agent **is situated** in a world (physical or virtual) where it receives inputs (perceptions) through sensors and influences the world through actuators. **Autonomy** implies that it operates without direct human intervention, making its own decisions about what actions to take [4]. **Flexible** action means that the agent can adapt to changing conditions and pursue its goals in a robust manner.

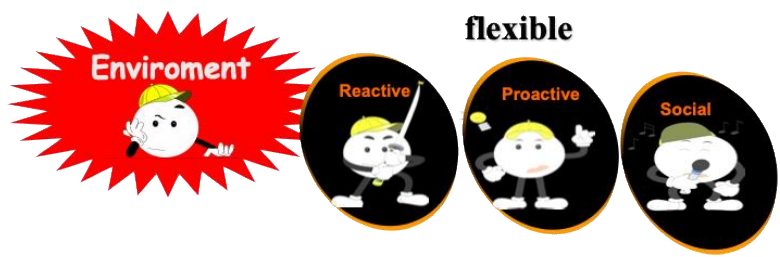

*Figure 1*. Definition of Wooldrige and Jennings Agent.

Wooldridge expanded that flexible and intelligent action involves properties such as **reactivity, proactivity, and social capability** [3]. These, along **with autonomy, are often cited as key properties of intelligent agents** [15]. **Autonomy** means that agents function independently, making decisions and executing actions without the need for direct human supervision; this autonomy allows them to perform tasks and adapt in real-time to changing environments. **Reactivity** implies that the agent perceives its environment and responds to changes in a timely manner. **Proactivity** means that it can take initiative—goal-oriented behavior—not only reacting to the environment, but also taking goal-directed actions to meet its goals. **Social capability** refers to an agent being able to interact with other agents (or humans)—communicating, cooperating, or competing as needed. These attributes distinguish an intelligent agent from a mere program acting in a predefined way. Importantly, these concepts were established in the mid-1990s, underscoring that the concept of AI agents is not new; researchers have long studied what makes an agent "intelligent."

To illustrate these ideas, we can visualize a simple model of an intelligent agent interacting with its environment (Fig. 2). The agent receives perceptions of the environment through sensors, processes these inputs (using its internal reasoning or decision-making mechanism), and then produces actions through actuators to change the environment. The **perceive-think-act cycle** is the core of any agent. For example, a thermostat agent perceives the current temperature, "decides" whether it matches the desired set point, and acts by turning the heater on or off. A more complex example is a robotic agent: it might perceive the world through cameras and distance sensors, reason about its goals (e.g., navigate to a location), and act on its propulsion system. In all cases, the notion of agency implies that the system continuously links perception to action in pursuit of goals.

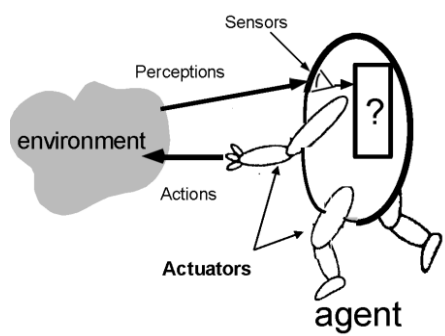

*Figure 2. Simple model of agent interacting with its environment.*

The box with the question mark if the Fig.2 is where we will incorporate the agent's intelligence, i.e. those AI techniques that allow us to determine from the agent's perceptions what actions to perform on the environment to achieve its objectives. These techniques will allow us to interpret its perceptions, to deliberate to determine which objectives to fulfill, to reason to determine which actions will allow it to achieve its objectives, etc. This involves the use of shape recognition techniques, natural language processing, planning, LLMs, etc.

It is worth noting that **Russell and Norvig's** AI textbook [5] identifies this agent-environment interaction as the fundamental concept of AI, even defining AI itself as the study of intelligent agents. By the late 1990s, the agent paradigm had become central to AI research, spawning subfields such as **autonomous agents and multi-agent systems**. This led to the creation of dedicated spaces such as the International Workshop on Agent Theories, Architectures, and Languages (**ATAL**), the International Conference on Autonomous Agents (**AGENTS**), the International Conference on Multi-Agent Systems (I**CMAS**), and their merger into the Autonomous Agents and Multiagent Systems Conference (**AAMAS**) starting in 2002. The professional societies [IFAAMAS](#) (The International Foundation for Autonomous Agents and Multiagent Systems) and [EURAMAS](#) (The European Association for Multi-Agent Systems), [FIPA](#) (The Foundation for Intelligent Physical Agents) and journals such as **Autonomous Agents and Multi-Agent Systems** (Springer) were established to advance the theory and practice of agents. The area of agents and multiagent systems has been present for many years at artificial intelligence conferences, in particular at the most prestigious IJCAI (International Joint Conference on Artificial Intelligence), **AAAI** (Association for the Advancement of Artificial Intelligence), and **ECAI** (European Conference on AI). In short, by the time current tech blogs started talking about "Agentic AI," the AI research community had already built a vibrant and evolving field around these concepts.

## Multi-Agent Systems

While a single intelligent agent can be powerful, many real-world problems require multiple cooperating (or competing) agents. The field of multi-agent systems (MAS) studies societies of interacting agents. A multi-agent system consists of multiple autonomous agents that interact—either collaboratively or competitively—to achieve individual or collective goals [4]. In such systems, no single agent has sufficient capabilities to achieve the overall objectives alone; agents must coordinate their actions and knowledge with others [4]. This interaction may involve communication, negotiation, cooperation on subtasks, or competition for shared resources, depending on the scenario.

By the early 2000s, MAS had matured as a distinct research area, addressing challenges of coordination, communication, and organization of multiple agents. Researchers developed standards and languages to enable interoperability between agents from different developers. For example, the Foundation for Intelligent Physical Agents (FIPA) began issuing standards for agent communication protocols in 1996 [11].

This area of research is based on a new paradigm of computing, computation as interaction [16]. In this paradigm, computation occurs through communication between computational entities; computing is an inherently social activity, not a solitary one, which gives rise to new ways of conceiving, designing, developing and managing computational systems.

When we talk about communication, we can't forget initiatives like the Knowledge Sharing Effort (**KSE**). This initiative was launched around 1990 by ARPA2(the Advanced Research Projects Association of the U.S. Department of Defense), with support from research organizations such as ASOFR, NSF, and NRI. Its main objective was to promote interoperability in the communication of knowledge between intelligent systems.

As a result of this project, it was concluded that effective communication between distributed agents requires three levels of language:
- Syntax: A language is required to represent knowledge in a formal and structured manner. http://www.cs.umbc.edu/kse/kif/
- Semantics: A language is required for defining ontology, which allows the meaning of the represented knowledge to be described. http://www.cs.umbc.edu/kse/ontology/
- Pragmatics: A language is required for communication between agents, focusing on how knowledge messages are exchanged and manipulated. http://www.cs.umbc.edu/kqml

The design of these languages was inspired by linguistic theories about speech acts [17, 18]:
- J.L. Austin (1962) – How to Do Things with Words
- J.R. Searle (1969) – Speech Acts

These works laid the groundwork for understanding the pragmatics of communication, that is, how words are used to perform actions.

Early agent communication languages were based on speech-act theory from philosophy of language (e.g., Austin's and Searle's work on speech acts) [17,18]. Concretely, languages like **KQML (Knowledge Query and Manipulation Language)** [19] and **FIPA-ACL** [20] were proposed to standardize how agents ask questions, inform each other, or negotiate. Underneath these, common formats like **KIF (Knowledge Interchange Format)** [21] for content syntax and languages, such as **ONTOLINGUA** [22], for defining shared ontologies for semantics were also developed. These efforts were the MAS community's attempt to ensure that heterogeneous agents could work together—a recognition that **social capability** (one of the properties of agents) required agreed-upon methods of communication.

In multi-agent systems, the agents that constitute them must interact to solve the problem for which they were designed; these interactions can be collaborative, when the agents are benevolent, or they can be competitive, when the agents are self-interested. This translates into the need for methods that allow agents to reach agreements or align their intentions – leading to concepts like **distributed consensus, teamwork, auctions for task allocation** (Agreement Computing [23]).

In response to this need to reach agreements, Agreement Technologies (AT) emerged [24]. These refer to computer systems in which autonomous software agents negotiate with each other, usually on behalf of people, to reach mutually acceptable agreements. An entity can choose whether to comply with an agreement or not, and must comply when there is an obligation derived from the current agreements. Autonomy, interaction, mobility and openness are relevant characteristics from a theoretical and practical perspective. Semantic alignment, negotiation, argumentation, normative systems, virtual organizations, learning and other technologies are part of a testing environment to define, specify and verify such systems. Security in execution is based on trust and reputation measures. These measures help agents determine with whom to interact and what terms and conditions to accept.

In Europe, for example, a large research program on **Agreement Technologies ([COST Action IC0801](#))** was launched in the late 2000s to investigate these issues. All of this underscores that, by 2010, the problems of coordinating multiple AI actors—from communication and cooperation to governance—were already actively studied and partially solved.

In summary, the AI community has long had clear definitions for what an **AI agent** is and what a **multi-agent system** is. These definitions revolve around autonomy, intentional action, and (for MAS) interaction among agents. Below we summarize the key characteristics of each concept:

- *Intelligent Agent:* an autonomous, flexible entity situated in an environment, with properties of autonomy, reactivity, proactiveness, and social ability.
- *Multi-Agent System:* a collection of autonomous agents interacting in a common environment, possibly cooperating or competing, requiring communication and coordination mechanisms.

The previous sections only briefly reviewed some of the key concepts in the field of agents/multi-agent systems, but from these concepts, it can be deduced that this is a well-established area in the field of AI.

In what follows, we focus on how AI agents are built—that is, the architectures that enable agents to exhibit intelligent behavior.

## Architectures of Intelligent Agents

Stuart Russell and Peter Norvig, in their book 'Artificial Intelligence: A Modern Approach' [5], a reference in teaching Artificial Intelligence, classify intelligent agents according to their ability to perceive, model, plan, and learn. The evolution goes from simple reactive agents to agents with autonomous learning capabilities, and they define five abstract agent architectures that we briefly describe below.

**Simple Reactive Agents:** They act directly based on current perceptions, using condition → action rules. They have no memory or internal model of the environment.

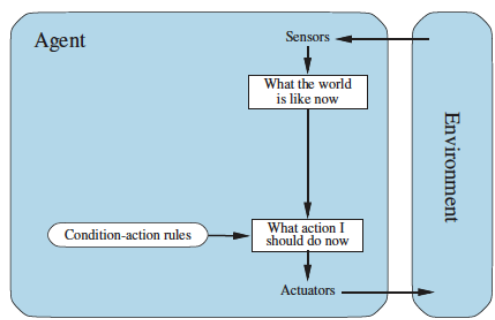

*Source: Artificial Intelligence: A Modern Approach [5]*

**Model-Based Agents** (memory): They maintain an internal model to represent the state of the environment. They update this model based on received perceptions.

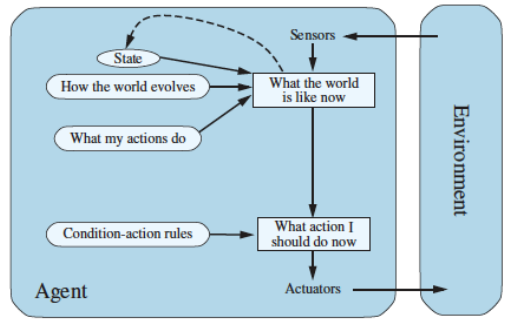

*Source: Artificial Intelligence: A Modern Approach [5]*

**Goal-Based Agents:** Select actions based on specific objectives. They use search and planning techniques to achieve these objectives.

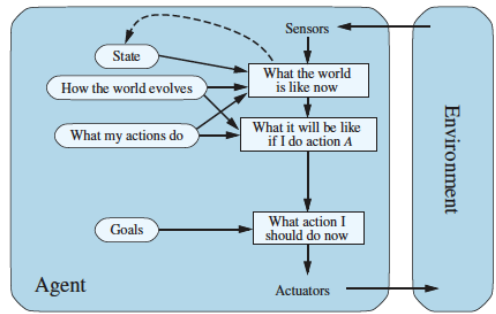

*Source: Artificial Intelligence: A Modern Approach [5]*

**Utility-Based Agents:** Generalize goal-based agents by using a utility function to evaluate different outcomes and select the most preferred one.

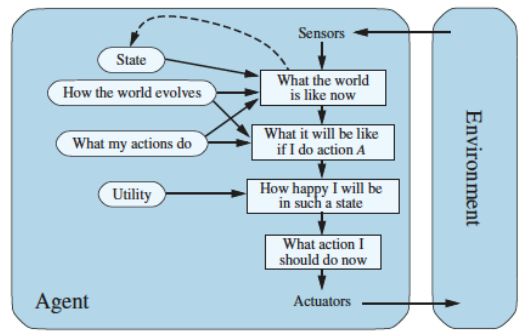

*Source: Artificial Intelligence: A Modern Approach [5]*

**Learning Agents:** Capable of improving their performance over time through experience. They include action, critique, and learning modules.

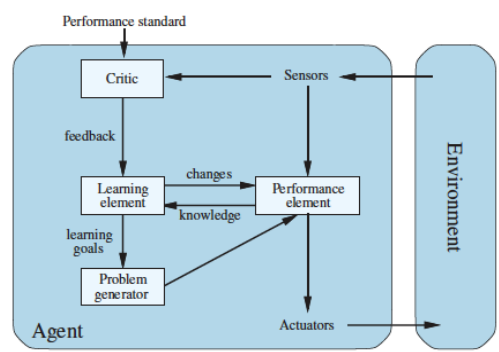

Designing an intelligent agent involves specifying how it processes information from perception to action. Over the years, AI researchers have proposed various agent architectures—underlying models that define an agent's internal structure and decision-making processes. They range from simple reactive systems to complex deliberative ones. Below, we review several of the major architectural paradigms, as understanding them will clarify how so-called "agentic" systems fit (or do not fit) with traditional approaches.

**Reactive (Reflex) Agents**: The simplest type of agent architecture is a **reactive agent**, also called a **simple reflex agent**. This agent selects actions based on current perception, ignoring perceptual history. It implements a direct mapping from situations to actions—essentially a set of condition-action rules. If the condition matches the current state of the environment, the agent executes the associated action; if no rules match, it does nothing. Reactive agents do **not perform explicit planning or model the world**; they operate in real time under if-then logic. This paradigm rose to prominence in the mid-1980s as a reaction to the limitations of pure symbolic planning in AI. **Rodney Brooks's subsumption architecture** [13] is an example. In this model, the agent's control system is built in layers of behavior (e.g., obstacle avoidance, wandering, goal seeking), where low-level reactive behaviors can override high-level ones when necessary. This was highly influential in behavior-based robotics, demonstrating that effective behaviors can emerge from networks of simple, reflex-like modules, without centralized symbolic reasoning. The advantage of reactive agents is their speed and robustness in dynamic environments (without complex deliberation to slow them down); the disadvantage is that purely reactive agents struggle with long-term strategic behaviors, as they lack an internal model of the world or explicit goals.

**Deliberative (Symbolic) Agents:** At the other end of the spectrum are deliberative agents, sometimes called **deductive reasoning agents** (especially in early literature). These agents maintain a symbolic model of the world and use symbolic reasoning (logic) to decide on actions. In a deliberative architecture, the agent has explicit representations of beliefs about the world and uses logical deduction or search to devise a plan of action that achieves its goals. This approach dates back to the early days of AI (1950s) and continued into the 1970s and 1980s in systems that attempted to prove theorems or derive plans from formal principles. A pure deductive agent might, for example, use automatic theorem proving to infer the best action given a formal goal and a formal knowledge base—a method that is correct by design but often computationally infeasible in complex, dynamic environments. By the late 1980s, it was recognized that purely symbolic agents were fragile and slow in practice; Real-life environments change faster than they can be replanned, and the framing problem (determining what changes and what stays the same after an action) makes simple logical approaches impractical. This realization led to the reactive agent movement mentioned earlier, as well as architectures that hybridize both extremes. Examples of this type of architecture are **AGENT0** and **PLACA**.

**Hybrid Agents: Hybrid architectures** attempt to combine the strengths of reactive and deliberative approaches [30]. A common pattern is a two (or three) layer design: a reactive layer handles low-level, time-critical responses to the environment, while a deliberative layer does higher-level planning and goal management. The two layers must be coordinated – for instance, the reactive layer might override plans to handle immediate threats, or conversely, the deliberative layer might suppress some reactive behaviors when pursuing a long-term plan. One famous early hybrid architecture was the **TouringMachines** architecture (Ferguson, 1992), which had layers for reactive skills, planning, and a supervisory layer to manage conflicts. Other examples include **InteRRaP** and the **3T (three-tier)** architectures. By the 1990s, hybrid agent design was the norm for complex robots and software agents: it acknowledged that **no single method is sufficient** for all aspects of intelligence. The hybrid approach recognizes the need for *reactivity* to handle real-time changes and *deliberation* to handle strategic reasoning. Many modern AI agents implicitly follow this layered approach, even if not explicitly labeled as such.

**Belief-Desire-Intention (BDI) Agents:** A particularly important architecture for intelligent agents is the **Belief–Desire–Intention (BDI) model**, which originates from the work of Michael Bratman[14] on human practical reasoning and was adapted for AI in the 1990s The BDI architecture provides a structured, theoretical framework for agent deliberation:

- **Beliefs (B):** the information state of the agent – what the agent believes about the world (which may be incomplete or incorrect).
- **Desires (D):** the objectives or situations that the agent would like to bring about (not all desires will be achievable or pursued).
- **Intentions (I):** the subset of desires that the agent has committed to achieving; in other words, its current goals or plan of action.

The BDI model defines a control loop whereby the agent continuously updates its beliefs, generates options (desires) based on those beliefs and current intentions, filters those desires to choose which to commit to (intentions), and then acts to fulfill the intentions. This maps closely to how humans reason about actions: we have many potential desires at any time, but we filter them based on priorities and feasibility into intentions, which we then try to execute. Crucially, a BDI agent doesn't blindly replan from scratch at every moment; it maintains **commitment** to intentions until certain conditions are met (goal achieved, goal impossible, or superseded by higher priority goal), providing a balance between proactive persistence and reactive adaptability.

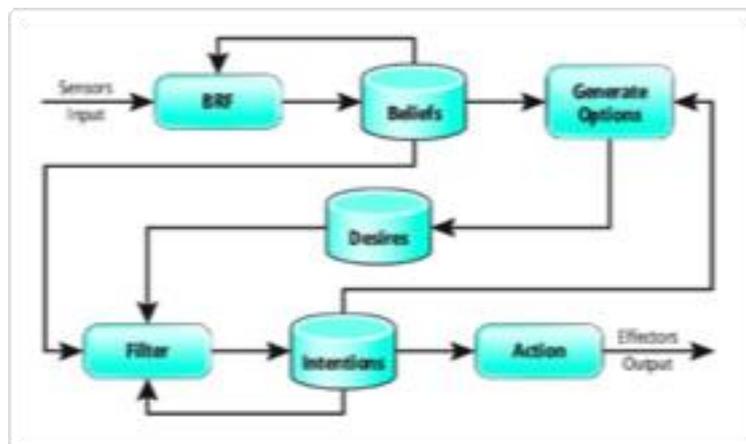

**Figure 3.** Schematic of a **BDI agent architecture**, showing how percepts from the environment are processed by the Belief Revision Function (BRF) to update the agent's **Beliefs**, which together with existing **Intentions** inform the generation of new **Options** (possible desires). A **Filter** then selects which desires to adopt as current **Intentions**, and the agent acts to achieve these intentions. This cycle of observing, updating beliefs, deliberating (options → intentions), and acting repeats continuously- *(Source: https://learn.microsoft.com/en-us/archive/msdn-magazine/2019/january/machine-learning-leveraging-the-beliefs-desires-intentions-agent-architecture)*

The BDI framework formalizes practical reasoning in agents, allowing them to balance reactivity (via perceptual updates) with goal-oriented behavior (via intentions). It has been used as the basis for many agent programming languages and platforms (e.g., PRS, JACK, JAM). The BDI model's emphasis on the agent's mental state (informational and motivational attitudes) connects with Dennett's intentional systems theory, providing a rigorous way to design agents that appear to have beliefs, desires, and intentions—in fact, in BDI, these are explicit data structures.

In addition to the architectures presented, there are numerous other architectures (event-driven agents, neural network agents, etc.), but those discussed illustrate the main paradigms. It is clear that, by the early 2000s, AI researchers had developed **a sophisticated set of tools for building agents**. Reactive architectures offered speed, deliberative architectures offered foresight, hybrid architectures attempted to combine the best of both, and BDI added philosophical clarity and practical structure for engagement strategies.

It is in this context that we consider the recent wave of "**Agentic AI**". When current commentary talks about agentic AI systems, we must ask: which of these established architectures (or combinations) are they actually using? Do today's "GPT-4-based autonomous agents" represent something fundamentally novel, or are they an implementation of, say, a BDI cycle with an LLM handling some of the decision-making? To explore this, we now examine what **Agentic AI** means in the context of current generative AI trends.

## From LLMs to "Agentic AI": The New Wave of Autonomous Agents

The year 2023 saw an explosion of interest in chaining and orchestrating large language models to perform goal-oriented autonomous tasks. The release of powerful LLMs (**GPT-3, GPT-4, GPTo, Gemini, Deepsek, Llama,** etc.) capable of reasoning and code generation led developers to create systems such as **AutoGPT, BabyAGI, LangChain, GrahpChain, AutoGen, CAMEL, ADK**, and many others, which have enabled the creation of agents and SMAs in which agents are LLMs capable of reasoning, planning, and communicating with each other using natural language. These LLM-based SMAs leverage the collective intelligence of multiple specialized agents to tackle complex tasks more efficiently than a single agent. A set of LLMs can assume specialized roles defined by prompts, collaborating through information exchange and coordination of actions in natural language, similar to human interaction. However, the success of LLMs in complex tasks requires considerable human intervention to guide text generation and decision-making. Therefore, autonomous coordination between LLMs poses several challenges, such as ensuring that agents cooperate in a way that aligns with human objectives.

This trend has been accompanied by the popularization of the term **"agentic AI."** But what exactly do people mean by this term?

In industrial usage, "Agentic AI" [8,9] generally refers to AI systems that can autonomously **decide and execute actions** to pursue a high-level goal, often by breaking the goal into subtasks and possibly coordinating multiple components or subagents. It implies a **degree of initiative (proactivity) and long-term planning** that goes beyond one-off responses or text generation. For example, an agentic AI assistant might not only respond to a user query, but also take proactive steps (e.g., schedule a meeting, send emails, execute transactions) without explicit prompts for each step. In essence, this describes an **intelligent agent** in the classic sense! The agent perceives a goal, iteratively decides "What subgoal or action should it take next to achieve it?", possibly invokes external tools or APIs, interacts with other agents, monitors progress, etc. The excitement around these capabilities in 2023–2024 is real—driven by **impressive demonstrations of LLM-based agents**—but conceptually, this can clearly be mapped to autonomous agent frameworks and concepts that have been around for decades.

It is revealing to see how modern sources distinguish "Agentic AI" vs "AI agents." According to an IBM article [7], "Agentic AI is the framework; AI agents are the building blocks within that framework." In other words, agentic AI refers to an overall system (perhaps an ecosystem of modules including one or more agents) designed to solve problems with minimal supervision, while an AI agent is a specific autonomous component that performs individual tasks within such a system. Softude [8] similarly describes an AI agent as "an Agentic AI building block… programmed to work autonomously," with perception, decision, and action loops, while Agentic AI refers to systems with a high degree of agency that "proactively set goals, make decisions, and act with minimal human intervention".

The key is that Agentic AI often involves not a single agent, but rather an **orchestrated set of agents and tools**, along with an architecture that allows them to dynamically coordinate. In cloud computing terms, an Agentic AI system might be thought of as a set or workflow (sometimes called an "**agent orchestration system**"), while an intelligent agent is an individual autonomous actor within it.

In [7] IBM's example scenario for Agentic AI is a **smart home energy system** that manages multiple device agents. The Agentic AI (the overall system) understands the homeowner's high-level goal (energy efficiency) and coordinates individual AI agents—the thermostat agent, the lighting agent, appliance agents—each with its own specific task and goal. This example corresponds to the concept of a **multi-agent system**: the smart home has multiple agents (thermostat, lights, etc.) that, in this case, cooperate. In fact, nothing conceptually exotic is introduced by calling it Agentic; it is essentially a smart multi-agent control system, a topic well covered in distributed AI research.

So why this new terminology? Part of it is likely due to marketing or the natural cycle of ideas being rediscovered by new communities. The resurgence of interest in "intelligent agents" among developers came after the advent of advanced LLMs, which provided a new set of tools for implementing agent behaviors (e.g., using LLMs to interpret goals and generate plans in natural language). To market this, terms like Agentic AI gained traction, perhaps to signal a shift from static AI models (which passively generate outputs) to AI systems that are active, goal-oriented participants (agents). The irony, however, is that the underlying notions of autonomy and multi-agent interaction are not new at all.

As previously discussed, we have also seen a proliferation of **open-source frameworks** and libraries to facilitate the construction of these LLM-based agents. In fact, by 2024–2025, dozens of new **LLM-based frameworks** have emerged, many released within a year. Examples include AutoGen, AgentScope, CreAI, CamelAI, ADK, OpenAI, LangChain, AutoGen, Hugging Face Transformers Agents, LangGraph, Agenta, Minichain, Auto-GPT, BabyAGI, and many more. These frameworks integrate LLM models that enable the development of LLM-based agents to reason, plan, and act in open environments. Their operation is based on three key capabilities:
- Reasoning: LLMs generate chains of thought to reason before acting.
- Memory: They integrate episodic and long-term memories (vector stores, logs) to contextualize decisions.
- Tool Use: Use external tools (calculators, browsers, APIs) through structured prompts.
- Providing abstractions for defining tools, managing instructions, and connecting multiple LLM-based agents.

Frameworks like **LangChain, AutoGen,** and **CrewAI** quickly gained tens of thousands of GitHub stars (gold bars), indicating strong developer interest, while many newer frameworks (to the right) entered "the scene" in late 2024 with growing traction (green bars show tech media mentions). This reflects a rapid **expansion of tools and libraries** for building autonomous intelligent agents, often centered around orchestrating LLMs with external tools. The trend underscores industry's excitement about agent-based AI solutions – but also risks fragmentation and repeated trials of ideas known in academic agent research.

The proliferation of frameworks suggests that many are tackling similar challenges: how to allow an AI agent to use tools (APIs, databases), how to maintain memory of past interactions, how to coordinate multiple agents or chain sub-tasks, etc. These are essentially implementations of what the agent research community would recognize as **planning, memory/knowledge base, and multi-agent coordination**. For instance, a common design pattern is an LLM-based "planner" agent that can spawn "executor" agents for sub-tasks, which then report back – reminiscent of hierarchical task planning or master-slave agent organizations from MAS literature.

Another example: the concept of giving an LLM agent access to tools (such as web search or calculators) and expecting it to decide when to use which one—this is analogous to an agent reasoning about its actions and choosing an appropriate effector or subroutine. Classically, this could be mapped to the agent's repertoire of actions and access to the environment. Today, researchers are inventing instructional techniques (e.g., "ReAct," which interweaves reasoning with tool use) to achieve these behaviors with LLMs, similar to previous work on agent planning systems that included both internal (reasoning) and external (executing an API call) actions.

Where new LLM-based systems really add value is in the **flexibility and knowledge** encapsulated by large, pre-trained models. The fact that an instance of GPT-4, Gemini, Llama, Deepsek, or any other model can parse an arbitrary instruction, decompose it into steps, and generate code or text to accomplish them is a leap in capability compared to, say, a hard-coded BDI agent that only knows a fixed library of plans. This has enabled much more general-purpose agents. For example, a well-designed LLM can serve as a universal interface between natural language objectives and formal actions, functioning as the agent's "brain" that reasons at a high level in an open-ended way. This wasn't possible with good reliability before deep learning, so the excitement around Agentic AI is justified: these systems can do things that previous agent systems struggled with, thanks to the power of LLMs.

The speed with which LLM-based agent and multi-agent systems development tools are being created is also a tremendously positive factor. During the evolution of the field of agents and multi-agent systems, the availability of development platforms (JADE, JACK, MADKit, ZEUS, Magentix, etc.) evolved very slowly, while today we have a wide range to choose from, and they continue to increase.

However, at a conceptual level, **agentic AI systems are still intelligent agents**. They must grapple with the same fundamental issues: autonomy (the system must decide and act without constant user intervention), reactivity (responding to changes or new information), proactivity (setting and pursuing intermediate goals), and social capability (if multiple agents are involved, they must communicate effectively). Similarly, a **multiagentic system** (a term occasionally used for LLM multi-agent scenarios) is, by definition, a **multi-agent system** and will encounter familiar phenomena such as emergent cooperative or competitive behavior, conflict resolution, and the need for communication protocols.

## Misuse of Terminology and Conceptual Conflation

From what we've discussed so far, we can address the crux of the matter: **the misuse of the terms "agentic" and "multiagentic"** in the **current discourse on AI**. Many recent articles and press releases use these words as if they denote a new technology, when in fact they are describing capabilities or designs of systems long examined under the banners of intelligent agents and multiagent systems. This conceptual confusion can be problematic for several reasons:

- **It obscures the wealth of existing research**. By failing to recognize that agentic AI is essentially about autonomous agents, new practitioners may overlook decades of literature on how to design and evaluate these systems. As a result, they may repeat past mistakes or rework previously solved problems. Indeed, one of the main motivations for this article is the concern that ignoring previous work will lead us to "reinvent the wheel" when applying agentic AI to new domains.

- It creates **terminological confusion**. Traditionally, in AI, the adjective "agentic" was rarely used; the term "agency" has been more prevalent; people simply talked about agent-based systems, autonomous agents, and so on. Now we see phrases like "agentic AI agents," which are redundant (literally, "agent-like AI agents"). The term "multiagentic" is even more problematic, as it attempts to adjective "multiagent." But "multiagent" already serves as an adjective (as in "multiagent system"). Saying "multiagent system" is an unnecessary neologism when "multiagent system" is clear and well-defined. This jargon can confuse newcomers and muddy communications, especially between industry and academia. For example, a software engineer might say, "We built a multiagent AI to do X," while an AI researcher would interpret that as "a multiagent system for X." It is to be expected that they mean the same thing, but divergent terminology can impede the transfer of knowledge.

- It can exaggerate novelty. When a new term is coined, it is sometimes implied that the concept itself is new. This can lead to inflated expectations or misplaced credit. Consider statements like "Agentic AI takes things a step further by enabling autonomous decision-making and task execution," a fair description, but one that has applied equally to inyelligent agents since the 1990s. Or statements like "Agentic AI refers to systems in which multiple AI agents collaborate... As opposed to a single inelligent agent, which operates in isolation," which basically describes a multi-agent system. Without historical context, without scientific rigor, readers may think these capabilities were invented by today's AI labs.

Let's be clear: **the term "agentic" is not a new concept in AI**, it has roots in the social and cognitive sciences and was later adopted (moderately) in AI, but the underlying ideas have been part of the AI agent parade for a long time. When 2023 technical papers talk about "Agentic AI," they are usually referring to what AI researchers would simply call an intelligent agent or an agent-based system. I believe the current use of the term "Agentic" is incorrect; when these terms are used, they are referring to the concept of an intelligent agent/multi-agent system, therefore that is the terminology that should be used. In other words, we should call a thing by its name: if it's an AI agent, call it an agent; if it's a multi-agent system, call it that.

To be clear, none of this is meant to diminish or downplay recent engineering achievements in implementing LLM-based agents with new AI tools. The advances are real, but they fit within an existing conceptual framework. By recognizing this, developers and researchers can greatly benefit from **the lessons already learned**. For example:

- The MAS community has studied emergent behaviors in multi-agent systems. Some of the first experiments with LLM-based multi-agent systems have observed surprising behaviors (positive or negative) when several LLM agents interact (e.g., getting stuck in loops of asking each other questions). Many of these could be understood using game-theoretic or coordination approaches developed for MAS.

- Issues of trust and safety in autonomous agents (such as an AI agent becoming uncontrollable or causing unintended harm) have long been discussed in the context of bounded autonomy and adjustable autonomy. Frameworks exist for incorporating constraints or ethical standards into agent decision-making. Rather than starting from scratch, new Agentic AI initiatives can adapt these frameworks to LLM-based agents.

- The importance of standards cannot be overstated. In the late 1990s, agent researchers realized that without standard interfaces, it is difficult to get agents from different sources to work together. In the current proliferation of agent frameworks, we also see fragmentation: one framework may not easily communicate with another. The AI industry would do well to remember the value of standards. Encouragingly, some initiatives are emerging, such as Anthropic's Model Context Protocol (MCP) in 2023 [12], which aims to standardize how AI assistants (agents) interact with external systems. This is a step in the right direction, but more collaboration is needed to avoid a situation where every agent framework is isolated from the others.

Another aspect that is confused in the media is the over-equation of the concept of an agent with LLMs. While the current enthusiasm stems largely from the capabilities of LLMs, the concept of an agent does not mandate the use of an LLM. An agent could use traditional rule-based logic, a reinforcement learning policy, or any combination of AI techniques. In fact, the best solutions will likely combine various AI techniques; for example, a **hybrid AI approach** where symbolic reasoning (for long-term planning or ensuring constraints) complements subsymbolic methods such as neural networks for perception or natural language processing. Some have termed this convergence **neurosymbolic AI**. Agent architectures are a natural place for such convergence: an agent could use a neural model to interpret a situation and a symbolic planner to decide on a sequence of actions, or an LLM to provide conversational capability to the agent. By reintegrating agent research with modern AI, we can achieve systems that truly exhibit robust intelligence.

# Toward Integration: Embracing the Agent Research Legacy in Modern AI

How can we move forward in a way that benefits from both new advances and established foundations? First, the AI community (both researchers and practitioners) should strive for **clear terminology** and conceptual consistency, always rooted in scientific rigor. If you are building a system where multiple LLM-based components interact, recognize that you have built a multi-agent system and consult the MAS literature for design guidance and potential pitfalls. If your application involves an autonomous decision-making cycle, consider it an intelligent agent and leverage familiar agent architectures (reactive, BDI, etc.) to structure it. In publications and discussions, using the standard terms (with perhaps a note indicating that they align with the popular term "Agentic AI") will facilitate interaction between experienced agent researchers and the new generation of developers excited about LLM-based agents.

Second, we should actively incorporate proven results from the agent field into new LLM-based agent frameworks. For example, the concept of an agent communication language could be very relevant if we start having multiple LLM-based agents communicating with each other. Currently, most inter-agent communication in LLM systems is done in ad-hoc natural language (literally, having one agent give directions to another in natural language). This is fine for prototyping, but as complexity increases, the need for a more structured exchange might be rediscovered (to avoid misunderstandings and reduce verbosity). Researchers in the 1990s already developed speech-act-based communication protocols; these could inspire more efficient prompting schemes or API-based message passing between agents. Recent work such as the Self-Referential Communication (SRC) protocol for LLM agents [25] or Agora ( A Scalable Communication Protocol for Networks of Large Language Models) [26] echoes these ideas, albeit couched in modern terms.

Similarly, MAS coordination and competition strategies (such as task assignment algorithms, consensus protocols, voting mechanisms, automatic negotiation, automatic argumentation, etc.) can be transferred to agent-based AI systems to manage multiple agents working on subproblems. If, for example, a team of AI agents is created to act as a virtual sales force (one finds leads, another writes emails, another schedules calls), a multi-agent resource allocation and scheduling problem arises. Instead of a naive trial-and-error approach, existing MAS coordination algorithms could be applied.

But let's not forget that in many other problematic domains, we will need to develop open multi-agent systems, and to do so, we will have to take into account the results and concepts already developed in the field of multi-agent systems, primarily standards, and agreement technologies to develop agent choreographies.

On the other hand, the agent research community must also update its tools by embracing the advantages of LLMs and modern machine learning. Classical agent architectures often struggled with knowledge acquisition and flexibility, areas where LLMs excel. A BDI agent with an LLM-based planner or LLM-based perception module could be much more powerful than a purely symbolic BDI agent. Integrating **learning** into agent architectures (so that agents improve with experience) has been a long-standing challenge. Current reinforcement learning and few-shot learning techniques may finally enable this in practice. In short, the **synergy between the wisdom of traditional agents and the new capabilities of AI** can push the frontiers of autonomous systems.

## Conclusion

"Agentic AI" has become a buzzword capturing the imagination of the tech world, but, as we have argued, it is essentially a **rebranding of established concepts of autonomous intelligent agents**, contextualized by recent advances in generative AI. The excitement around LLM-based autonomous agents is justified by their impressive new capabilities, but it should not lead us to ignore the vast body of knowledge about agents and multi-agent systems developed over decades. The term "agentic" itself, borrowed from the social sciences, highlights intentional and goal-oriented behavior—precisely the focus of research on intelligent agents since the 1990s. Similarly, "multiagentic" scenarios are simply multi-agent systems under another name. Using these terms as if they denote novel ideas risks generating confusion and repeating past work and mistakes.

By reviewing the theoretical foundations (Bandura's agency, Dennett's intentional systems) and the AI community's prior achievements (definitions, properties, architectures, and frameworks for autonomous agents), we emphasize that **the wheel has already been invented** when it comes to autonomous agents. Fortunately, it's a pretty good wheel—one that can move forward with the momentum of generative AI. We encourage current practitioners to build on prior research. In practical terms, this involves adopting correct terminology, consulting the agent / multi-agent systems literature for guidance, and integrating proven design principles (such as communication standards, agent architectures, agreement technologies, ..) into new AI systems.

Conversely, we also recognize that new technology often requires adaptations of old ideas. The agent community must seize the opportunities offered by LLMs and deep learning to enhance agents' reasoning and learning capabilities. The result of combining these approaches will be truly **advanced autonomous agents** that are robust, adaptable, and capable of tackling complex real-world tasks. Indeed, as one vision suggests, the future could lie in **neurosymbolic agents**—systems that combine perception and language understanding powered by generative AI with symbolic reasoning and planning. In my view, this integration of agents, generative AI, soft AI, and responsible AI (RAI) is the path to **neurosymbolic AI or hybrid AI**, which is the next step in the evolution of AI and also likely to the near future of embodied AI.

In conclusion, the field of AI must be careful not to lose sight of its own history amid the current enthusiasm. The concepts of agents and multi-agent systems remain as relevant in 2025 as they were in 1995, even if we implement them today with far more powerful tools. Instead of coining ambiguous terms like "agentic AI," we should call these systems by their name and build on the foundations already established. By doing so, we not only give credit where credit is due but also accelerate progress—by avoiding retreading familiar ground and focusing on genuine innovations. The wheel of intelligent agents doesn't need reinventing; it needs refinement and turbocharging with the engines of generative AI. The path ahead for autonomous agents is exciting, and by uniting past knowledge with present technology, we can ensure that it builds on an already charted route, taking us to destinations beyond what previous generations of AI were able to reach.